\documentstyle[11pt,paspconf]{article}

\begin{document}

\title{Where's the Disk?: LBV bubbles and Aspherical Fast Winds}

\author{Adam Frank} 
\affil{Department of Physics and Astronomy,
University of Rochester, Rochester, NY 14627-0171;\\ 
email: afrank@alethea.pas.rochester.edu}

\begin{abstract}
Previous studies have explained the shapes of LBV nebulae, such as
$\eta$ Car, by invoking the interaction of an isotropic fast wind with a
previously deposited, slow {\it aspherical} wind (a "slow torus").  
In this work
I focus on the opposite scenario where an aspherical fast wind expands into a
previously deposited {\it isotropic} slow wind. Using high resolution
hydrodynamic simulations which include the effects of radiative cooling I have
completed a series of numerical experiments to test if and how aspherical fast
winds effects wind blown bubble morphologies.  The
simulations demonstrate that aspherical fast winds can produce strongly bipolar
outflows and recover some important aspects
of LBV bubbles which the previous models can not.
\end{abstract}



\section{Introduction}

In just a few years the HST has transformed our
understanding of the massive unstable stars know as Luminous Blue
Variables (LBVs).  Recent observations have revealed a number of LBVs
or LBV candidates to be surrounded by extended {\it aspherical}
outflows. The most extraordinary of these is the markedly bipolar
nebula surrounding $\eta$ Carinae (``the homunculus'':
Hester {\it et al.} 1991,Ebbets {\it et al.} 1993, Humphreys \& Davidson 1994).  Other LBVs show
nebulae with varying degrees of asphericity from elliptical
(R127, Nota {\it et al.} 1995) to strongly bipolar (which we define though the
presence of an equatorial waist:
HR Carinae; Nota {\it et al.} 1995; Weis {\it et al.} 1996).

These bipolar morphologies are quite similar to what has been observed in
Planetary Nebulae (PNe) which arise from low mass stars
(Manchado {\it et al.} 1996). The aspherical shapes of PNe
have been successfully explained through a scenario termed the ``Generalized
Interacting Stellar Winds'' model (GISW,
Frank \&\ Mellema 1994) where an
isotropic fast wind from the central star (a proto-white dwarf) expands
into an aspherical (toroidal) slow wind ejected by the star in its
previous incarnation as a Asymptotic Red Giant. High densities in the
equatorial plane of the AGB outflow constrain the expansion of the fast
wind. 
The expanding shock wave which results from the wind/wind interaction
quickly assumes a
elliptical prolate geometry.  If the ratio of mass density between
the equator and pole (a
parameter we call q, $q_{\rho} = \rho_e/\rho_p$) is high enough, then the
elliptical bubble eventually develops a waist and becomes bipolar.

The similarity of PNe and LBV nebulae has led to the suggestion that
both families of objects are shaped in similar ways.  In Frank, Balick,
\& Davidson 1995 (hereafter: FBD) a GISW model for $\eta$ Car was
presented in which a spherical ``outburst'' wind expelled during the
$\approx$ 1840 outburst expanded into a toroidal ``pre-outburst''
wind.  FBD showed that the resulting bipolar outflow could recover both
the gross morphology and kinematics of the Homunculus.  Nota {\it et
al.} 1995 (hereafter NLCS) used a similar model for other LBV nebulae
presenting a unified picture of the development of LBV outflows.  More
recently Mac Low, Langer \& Garcia-Segura 1996 (hereafter MLG)
presented a model which also relied on the GISW scenario but which
changed the order of importance of the winds.  Using the Wind Compressed
Disk model of Bjorkman \& Cassinelli 1992 MLG showed that a strong
equator to pole density contrast would likely form {\it during the
outburst}.  In MLG's model it is the post-outburst wind which
``inflates'' the bipolar bubble via its interaction with the toroidal
outburst wind.

While all these models have demonstrated the potential efficacy of the
GISW scenario via a reliance on a slow torus
they are are troubling in their mutual inconsistency.  Specifically
the question - ``Where is the disk (torus)?'' - must be answered.  Does the
torus form during the LBV eruption as in MLG or does it form during the 
pre-eruption wind as in FBD and NLCS?  

Stepping back further one can also ask if a disk is needed at all?  The
latter question arises from consideration of new HST images of $\eta$
Car (Morse {\it et al.} 96) which reveal the equatorial
``disk'' to be so highly fragmented
that it may be more reasonable to consider it a ``skirt''
of individual clumps of ejecta rather than a azimuthally continuous structure.
 This
point is crucial since an equatorial spray of isolated bullets can not
hydrodynamically constrain an isotropic stellar wind to form a bipolar
outflow.  Thus we are led to consider an alternative model
to that proposed by FBD, NLCS and MGL.   Here we further generalize the
GISW model by turning that scenario on its head.  It what follows we
consider the case of an {\it aspherical fast wind} interacting with a
{\it isotropic slow wind}.  We imagine a fast wind ejected with higher
velocity along the poles than along the equator.  The question we wish
to answer is: can such a wind, expanding into a isotropic environment,
account for the shapes of LBV nebulae.

Theoretical models admit the possibility of aspherical fast winds in
massive stars (Lamers \& Pauldrach 1991).  More importantly, there is
direct evidence for asphericty in LBV winds. Observations of the wind
of AG Carinae (Leitherer {\it et al.} 1994) imply a pattern of
densities and velocities from pole to equator much like that described
in Lamers \& Pauldrach 1991.  Finally we note that it is worthwhile to
pursue this kind of investigation simply because it has not been
done before.  The GISW model and its variations has been very successful
in accounting for a variety of bipolar outflow phenomena 
(Blondin \& Lundqvist 1993; Frank, Balick \&
Livio 1996; Frank \& Mellema
1996,).  Since the effect of aspherical fast winds has yet to be
investigated the potential of finding useful results is high which is
argument enough for a detailed study.

\section{Computational Methods and Initial Conditions}

The details of the computational method and initial conditions can be
found elsewhere (Frank, Ryu \& Davidson 1997).  Here we present only a
brief overview.  We model the gasdynamic interactions via the Euler
equations with a radiation
loss `sink' term in the energy equation.

Our numerical experiments are designed to study the evolution of
wind-blown bubbles driven by aspherical fast winds.  The environment is
always assumed to be characteristic of a previously deposited spherically
symmetric ``pre-outburst'' wind which we denote as wind 0 with mass loss
rate $\dot M_0$ and velocity $V_0$.  
For the driving ``fast'' or ``outburst'' wind, which we denote
as wind 1, we need a formalism for setting the latitudinal ($\theta$)
variations in the wind properties i.e. $\dot M_1 = \dot M_1 (\theta)$
and $V_1 = V_1(\theta)$.  We note that since we wish to drive prolate
bipolar bubbles we always assume that the velocity at the poles is
larger that at the equator.  We have also explore models with
a ``post-outburst'' wind (denoted as wind 2)

We have chosen to explore different scenarios for the pole 
to equator variation in
wind parameters.  Each scenario is based on assuming a different
quantity remain constant across the face of the star.  They 
are: Constant Momentum Input ($\dot \Pi = \dot M_1 V_1$ =
Const); Constant Energy  Input ($\dot E = {1 \over 2} \dot M_1 {V_1}^2$
= Const); Constant density ($\rho_1 = $ Const).

If we choose our fiducial values for the density and velocity at the
equator $(\rho_{1e},V_{1e})$, the latitudinal variation can
be expressed as powers of an ad hoc function $f(\theta)$ which produces
a smooth variation in $V_1$ and $\rho_1$ from equator to pole.  Using
the formalism described above we have carried out three sets of
numerical experiments varying the ratio of mass loss rates
in the successive winds between each set.  Within the first
two sets we performed
three simulations with $\dot M_1 (\theta)$
and $V_1(\theta)$
corresponding to the scenarios discussed above.  In the final set
only the $\dot \Pi = Const$ case was used and the equator to pole
velocity contrast $q_v = V_{1e}/V_{1p}$ was varied.  The initial
conditions for each of our 9 simulations are shown in table 1.

\begin {table}[hbta]
\caption {Initial Conditions For Runs A - H}

\begin {center}
\begin {tabular} {lllllllll} \hline
{run}  & {${\dot M}_{\rm 0}$} & {$V_{\rm 0}$} & {${\dot M}_{\rm 1e}$} 
& {${V}_{\rm 1e}$} &  {${\dot M}_{\rm 2}$} 
& {${V}_{\rm 2}$} &   {$q_v$} \\
\hline
A-C  &  $1 \times 10^{-4}$ & 100  & $1 \times 10^{-4}$ & 150 & 
NA & NA & $0.2$ \\
D-F  &  $1 \times 10^{-6}$ & 100  & $1 \times 10^{-4}$ & 150 & 
NA & NA & $0.2$ \\
G  &  $1 \times 10^{-6}$ & 100  & $1 \times 10^{-4}$ & 150 & 
$1 \times 10^{-6}$ & 1400 & $0.3$ \\
H  &  $1 \times 10^{-6}$ & 100  & $1 \times 10^{-4}$ & 150 & 
$1 \times 10^{-6}$ & 1400 & $0.14$ \\
I  &  $1 \times 10^{-6}$ & 100  & $1 \times 10^{-4}$ & 150 & 
$1 \times 10^{-6}$ & 1400 & $0.1$ \\

\end {tabular}
\end {center}
\end {table}
 
\section{Results}
\subsection{2 Wind Models with $\dot M_{0} = \dot M_{1e}$}
In Experiment 1
we examined the interaction between two winds of with the same 
mass loss rate.
These simulations are performed to give us a baseline on the
gas-dynamic flow pattern.  The results of these simulations are shown
in the top row of Fig 1 where we present grayscale maps of $Log_{10} (\rho)$
for models A ($\dot \Pi$ =
Const), B ($\dot E$ = Const), and C
($\rho$ = Const).  Fig 1 shows that in all three scenarios the shell
of swept-up pre-outburst material becomes significantly aspherical
due to the
aspherical driving force of the outburst wind.   Notice also that the
bubbles all develop a ``waist'' - the observational signature
of a bipolar, rather than elliptical, configuration.  
Model C has the highest global
outburst wind density demonstrating that the degree of bipolarity
depends on relative densities between the aspherical outburst
and spherically symmetric pre-outburst winds.   

\subsection{2 Wind Models with $\dot M_{1e} > \dot M_0$} 
As Langer {\it et
al.} 1994 have demonstrated the LBV
outburst phase is likely to involve an increase in mass loss over the
pre-outburst wind.  Thus we have the case of a ``heavy'' wind expanding
into a light one.  To explore this situation we have run a second set
of experiments, runs D, E, \& F, where $\dot M_{1e} = 100 \dot M_0$.  
The results of these simulations are shown in the second row of Fig 1.

The bubbles formed in these runs are more strongly bipolar than
those in the previous experiment.  The reason for this is the relatively
small effect of the ambient medium in decelerating the massive winds
(deceleration does however occur).  Runs E and F show similar
morphologies as was the case in the first set of experiments. 
Similarly the $\rho =
Const$ case again produces the most bipolar configuration because of the
density is high across the face of the star. Fig 1 demonstrates the
principle conclusion of our first two sets of experiments. {\it
An aspherical stellar wind can drive an aspherical bubble}.

\subsection{3 Wind Models} In runs G, H, and I we have performed
simulations similar to run D. In the new
simulations the outburst wind lasted only 30 years. Afterwards
a post-outburst wind was
driven into the grid.  The characteristics of the post-outburst wind
were meant to qualitatively 
mimic the conditions currently observed in $\eta$ Carinae. 
We used a relatively low mass rate, high velocity
post-outburst wind i.e.  $\dot M_2 < \dot M_{1e}$ and $V_2 > V_{1e}$.  Thus in
these simulations (each of which has a different equator to pole velocity
contrast) we are interested the effect of the post-outburst wind 
on a bipolar bubble created by the previously ejected dense
aspherical outburst wind.

In the last row of 
Fig 1 we present $log_{10} (\rho)$ maps for all three simulations
in these experiments.  Runs G,H, and I with $q_v = 1/3, ~1/7$
and $1/10$ all produce strong bipolar
morphologies which develop without the need for a slow-moving disk.
Comparison between the simulations shows that decreasing $q_v$ produces
stronger bipolar morphologies.   It is
worth noting that if the expansion of the bubble were ballistic we
would expect the shape of the bubbles to scale with $q_v$.
Since this is not the case the bubbles must experience 
significant hydrodynamic
shaping (i.e. what you put in is not what you get out).
Some part of the shaping is due to
deceleration of the outburst wind via the previously ejected
material.  The post-outburst wind however also contributes by
accelerating the the outburst material.  As the bubble evolves
this acceleration will have its greatest effect near the equator where
the outburst wind has been most strongly decelerated.  Thus the action
of the post-outburst wind will be to drive the bubble towards a more
spherical configuration as system evolves.

\section{Discussion and Conclusions} 
Our simulations
demonstrate that bipolar wind blown bubbles can result purely from the
action of an aspherical fast wind.  In previous studies of LBV bubbles
(FBD, NLCS, MLG) it has been assumed that a slow moving torus or disk
of gas must exist first for a successive spherical fast wind to
shape into a bipolar configuration.  Our results indicate that
the properties of LBV bubbles may not require such a torus to form
either before (FBD, NLCS) or during (MLG) the outburst. 

Thus we are faced with an abundance of models to explain the same 
phenomena. Resolving the issue of if a disk is needed may
require observations of the angular profiles of mass and momentum in 
LBV shells (such an approach has been successfully used for YSOs bipolar
bubbles; Masson \& Chernin 1992) 
\bigskip
{\center{\bf Acknowledgments}}
Support for this work was provided
by NASA grant HS-01070.01-94A
from the Space Telescope Science Institute, which is operated by
AURA Inc under NASA contract NASA-26555.  Additional support came from the
Minnesota Supercomputer Institute.

\begin{question}{Dr.\ Ignace}
Aside from $\eta$ Car are you suggesting that the brightness enhancements
around, say, AG Car and R127, are from dense polar caps instead of a dense
waist and as a discriminant would you predict the caps to be expanding 
faster than the equator.
\end{question}
\begin{answer}{Frank}
The waist may still appear brighter because emission is a density
squared process but you are right that the kinematical pattern for
these kinds of bubbles may be different from the slow torus version
of the bipolar bubble.
\end{answer}
\begin{question}{Dr.\ Owocki}
It is interesting that you get the best agreement with the shape of $\eta$
Car when you use an constant density wind.  Radiation driven winds from
rotating stars with gravity darkening predict higher densities at the pole
so I suggest you explore that case two.
\end{question}
\begin{answer}{Frank}
In such a case an even more bipolar bubble would be expected since you
are increasing the ram pressure at the pole.  I will include that in my 
future models.
\end{answer}
\begin{question}{Dr. Langer}
I wonder if a fast dense outburst wind is compatible with the models:
firstly if the outbursts are related to surface instabilities the escape
velocity goes to zero and the winds would be slow.  There would also be an
energy problem: the kinetic flux might overwhelm the stellar luminosity.
\end{question}
\begin{answer}{Frank}
Well fast and slow are relative concepts.  All I am asking is that
an aspherical outburst wind drive the shape of the nebula.  I use velocities as
low as 100 km/s in my models.  I don't intrinsically
need $V_{wind}$ to be an order of magnitude higher. Also the pre-outburst wind
may be relatively fast and have left a shocked bubble behind which will still
provide an isotropic back-pressure.
\end{answer}
\begin{question}{Dr.\ Schulte-Ladbeck}
What was the geometry of the post-outburst wind?  Also using 
spectropolarimity we have shown that the winds of R127, AG Car \& HR Car are 
axisymmetric today which indicates the post-eruption wind is probably
asymmetric as well
\end{question}
\begin{answer}{Frank}
That is a good point. I chose a spherical post-outburst wind for simplicity.  
Making it aspherical would however not change the qualitative effects 
seen in these models.
\end{answer}

\end{document}